\documentclass[prb,aps,twocolumn,amssymb]{revtex4}
\usepackage{graphicx,amsmath,amssymb,mathrsfs,dsfont,times}

\renewcommand{\b}[1]{\mathbf{#1}}
\renewcommand{\c}[1]{\mathcal{#1}}
\renewcommand{\u}{\uparrow}
\renewcommand{\d}{\downarrow}
\newcommand{\bsigma}{\boldsymbol{\sigma}}

\newcommand{\tr}{\mathop{\mathrm{tr}}}

\newcommand{\sh}{{\textstyle{\frac{1}{2}}}}

\newcommand{\bigS}{\mathcal{S}}

\newcommand{\EF}{\varepsilon_F}

\begin{document}

\title{Spin Aharonov-Bohm effect and topological spin transistor}

\author{Joseph Maciejko$^{1,2}$, Eun-Ah Kim$^3$, and Xiao-Liang
Qi$^{1,2}$}

\affiliation{$^1$Department of Physics, Stanford University,
Stanford,
    CA 94305, USA\\
    $^2$Stanford Institute for Materials and Energy Sciences, SLAC
    National Accelerator Laboratory, Menlo Park, CA 94025, USA\\
    $^3$Department of Physics, Cornell University, Ithaca, NY 14853}

    \date{\today}
\begin{abstract}
Ever since its discovery, the electron spin has only been measured
or manipulated through the application of an electromagnetic force
acting on the associated magnetic moment. In this work, we propose
a spin Aharonov-Bohm effect in which the electron spin is
controlled by a magnetic flux while no electromagnetic field is
acting on the electron. Such a nonlocal spin manipulation is
realized in an Aharonov-Bohm ring made from the recently
discovered quantum spin Hall insulator, by taking advantage of the
defining property of the quantum spin Hall edge states: the
one-to-one correspondence between spin polarization and direction
of propagation. The proposed setup can be used to realize a new
spintronics device, the topological spin transistor, in which the
spin rotation is completely controlled by a magnetic flux of
$hc/2e$, independently of the details of the sample.
\end{abstract}

\maketitle

\section{Introduction}

The spin of the electron is one of the most fundamental quantum
mechanical degrees of freedom in Nature. Historically, the
discovery of the electron spin helped to lay the foundation of
relativistic quantum mechanics. In recent years, the electron spin
has been proposed as a possible alternate state variable for the
next generation of computers, which led to extensive efforts
towards achieving control and manipulation of the electron spin, a
field known as spintronics \cite{zutic2004}. Despite the great
variety of currently used or theoretically proposed means of
manipulating the electron spin, a feature common to all of them is
that they all make use of the {\em classical} electromagnetic
force or torque acting {\em locally} on the magnetic moment
associated with the spin.

On the other hand, it is known that due to the Aharonov-Bohm (AB)
effect \cite{aharonov1959}, electrons in a ring can be affected in
a purely {\em quantum mechanical} and {\em nonlocal} way by the
flux enclosed by the ring even though no magnetic field
--- hence no classical force --- is acting on them. This effect
could be termed `charge AB effect', as it relies only on the
electron carrying an electric charge. This observation leads
naturally to the question of whether it is possible to observe a
`spin AB effect' which would enable one to manipulate the electron
spin in a purely nonlocal and quantum mechanical way, without any
classical force or torque acting locally on the spin magnetic
moment.

In this work, we show that the spin AB effect is indeed possible by
making use of the edge states of the recently discovered quantum
spin Hall (QSH) insulators. In recent years, the QSH insulator state
has been proposed in several different materials
\cite{kane2005,bernevig2006a,Murakami2006,bernevig2006d,Liu2008,Shitade2009}.
In particular, this topologically nontrivial state of matter has
been recently predicted \cite{bernevig2006d} and realized
experimentally \cite{koenig2007b,Roth2009,Buttiker2009} in HgTe
quantum wells (QWs). The QSH insulator is invariant under time
reversal (TR), has a charge excitation gap in the bulk, but has
topologically protected gapless edge states that lie inside the bulk
insulating gap. These edge states have a distinct helical property:
two states with opposite spin polarization counterpropagate at a
given edge \cite{kane2005a,wu2006,xu2006}. The edge states come in
Kramers doublets, and TR symmetry ensures the crossing of their
energy levels at TR invariant points in the Brillouin zone. Because
of this level crossing, the spectrum of a QSH insulator cannot be
adiabatically deformed into that of a topologically trivial
insulator without closing the bulk gap. The helicity of the QSH edge
states is the decisive property which allows the spin AB effect to
exist: the perfect correlation between spin orientation and
direction of propagation allows the transmutation of a usual charge
AB effect into a spin AB effect, as will be explained in detail
below.

\begin{figure}
\begin{center}
\includegraphics[width=3.5in]{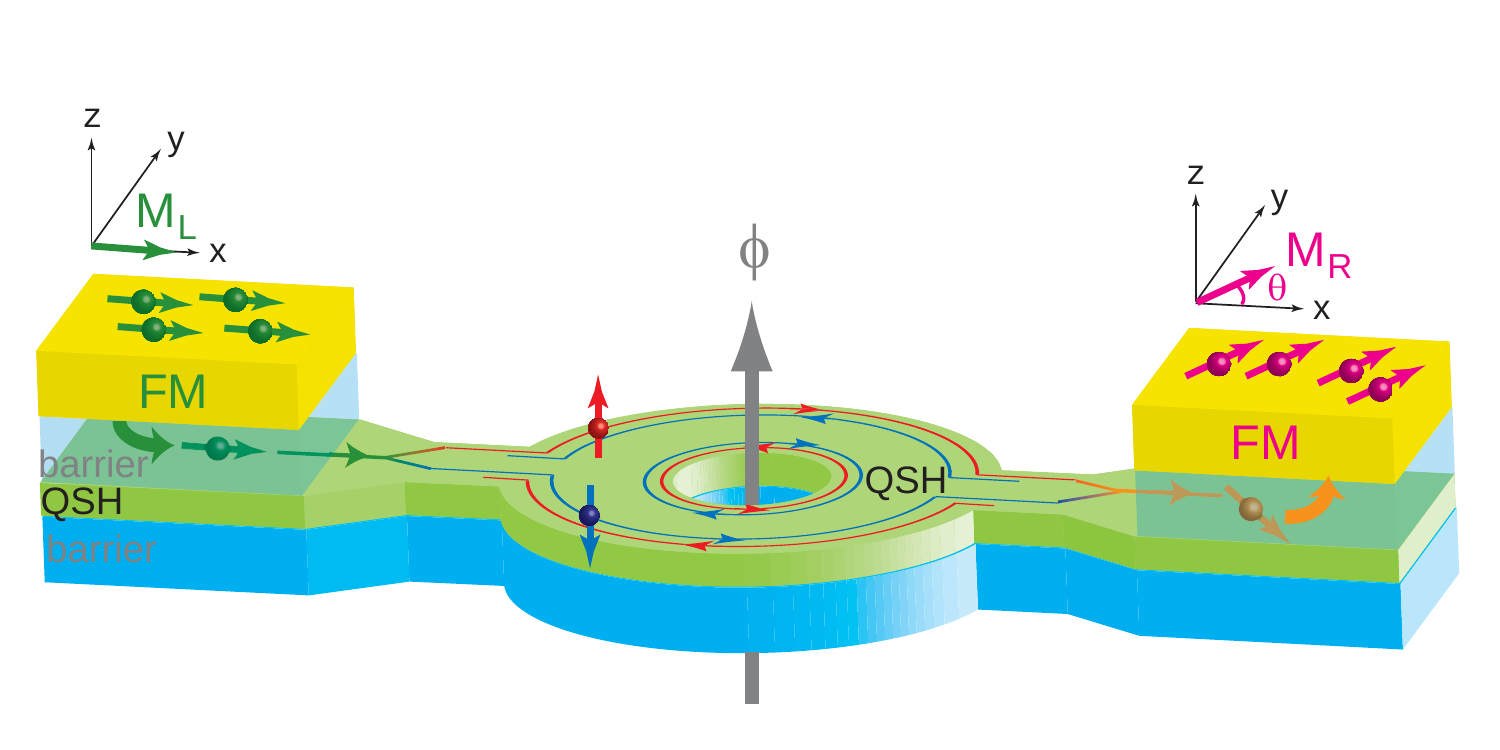}
\end{center}
\caption{{\bf Schematic picture of the spin AB effect.} A ring of
QSH insulator threaded by a magnetic flux $\phi$ is connected to two
magnetic leads. Spin polarized electrons injected from the left
lead enter the QSH region as a superposition of spin
up and down states. The spin up (down) state can only propagate along
the top (bottom) edge of the QSH ring, and the two spin states thus acquire an AB
phase difference proportional to $\phi$. Consequently, upon exiting
the QSH region the two edge states recombine into a state with
spin rotated with respect to the injected direction. The magnetization direction
of the right lead generally differs from that of the left lead by an
angle $\theta$. The two-terminal conductance $G=G(\phi,\theta)$ of
the device depends on the relative angle between the spin
polarization of the outgoing state and that of the right lead.
%A layer of FM material with magnetization $\b{M}_L$ along the $x$
%direction is deposited on top of the left ($L$) electrode, and
%injects a spin-polarized current into the underlying QSH region with
%electrons polarized along the $x$ direction (green arrows). Upon
%entering the QSH region (green), the $x$-polarized injected
%electrons split into two right-moving $z$-polarized QSH helical edge
%channels, with opposite spins propagating along opposite edges. A
%magnetic flux $\phi$ threading the center of the QSH region causes
%the electrons on different edges to acquire an AB phase difference
%$\varphi$ proportional to $\phi$. Upon exiting the QSH region, the
%helical electrons recombine into a state with rotated spin (orange
%arrow) where the rotation angle is dictated by $\varphi$. The
%two-terminal conductance $G(\phi,\theta)$ depends on the degree of
%overlap of this flux-rotated spin state with the spin state of
%polarized electrons (purple arrows) in the right FM layer with
%magnetization $\b{M}_R$ through which the current is collected.
}\label{fig:2terminal}
\end{figure}

The mechanism we propose to realize the spin AB effect is
illustrated in Fig. 1. Consider a two-terminal device consisting
of a bounded QSH insulator region pierced by a hole which is
threaded by a magnetic flux $\phi$. If the edge electrons
propagating clockwise have their spin pointing out-of-plane along
$z$ (spin up $\left|\uparrow\rangle\right.$, red trajectory), due
to TR symmetry the electrons propagating counterclockwise must
have opposite spin along $-z$ (spin down
$\left|\downarrow\rangle\right.$, blue trajectory). If we inject
electrons spin-polarized along the $x$ direction
$\left|\rightarrow\rangle\right.=\frac{1}{\sqrt{2}}
(\left|\uparrow\rangle\right.+\left|\downarrow\rangle\right.)$
from a ferromagnetic (FM) lead on the left, the electron beam will
be split coherently upon entering the QSH region at the left
junction into a $\left|\uparrow\rangle\right.$ beam propagating
along the top edge and a $\left|\downarrow\rangle\right.$ beam
propagating along the bottom edge. When the electron beams are
recombined on the right side of the ring, the electrons along top
and bottom edges will acquire a phase difference of
$\varphi=2\pi\phi/\phi_0$ due to the AB effect, where
$\phi_0=hc/e$ is the flux quantum. Consequently, the output state
is given by $\frac{1}{\sqrt{2}}\bigl(\left|\uparrow\rangle\right.
+e^{-i\varphi}\left|\downarrow\rangle\right.\bigr)$, such that the
electron spin is rotated by an angle $\varphi$ in the $xy$ plane.
The magnetic flux being confined to the hole in the device (Fig.
1), the electromagnetic fields are zero in the region where the
electrons propagate, and the spin is rotated by a purely quantum
mechanical Berry phase effect. In particular, for collinear FM
leads ($\theta=0$ in Fig. 1), one expects the conductance to be
maximal for $\phi=0\,(\mathrm{mod}\,\phi_0)$ and minimal for
$\phi=\frac{1}{2}\phi_0\,(\mathrm{mod}\,\phi_0)$, thus realizing a
``topological" spin transistor (see Fig. 3 c). This effect is
topological in the sense that the spin is always rotated by one
cycle for each period of flux $\phi_0$, regardless of the details
of the device, such as the size of the system or the shape of the
ring.

\section{Phenomenological scattering matrix analysis}

Before considering any microscopic model of transport in a QSH
system, generic features of two-terminal transport in the device
of Fig. 1 that depend only on symmetry considerations can be
extracted from a simple phenomenological scattering matrix or
$S$-matrix analysis \cite{stone1988}. The left and right junctions
are each described by a scattering matrix $S_L$ and $S_R$,
respectively (e.g. Fig. 2a for the left junction). Considering the
left junction first, $S_L$ consists of four submatrices
$t_L,t'_L,r_L,r'_L$ which correspond respectively to transmission
from left to right, transmission from right to left, reflection
from the left, and reflection from the right. One can define
similar submatrices for $S_R$. We wish to obtain an effective
$S$-matrix $S$ (see equation (\ref{trans})) for the whole device,
by combining the $S$-matrices of the junctions together with the
$S$-matrix for the central QSH region. Inside the QSH region, the
AB effect is described by the matrix $\Phi\equiv
e^{-i\varphi\sigma_z/2}$ where $\sigma_x,\sigma_y,\sigma_z$ are
the three Pauli matrices. In addition to the geometric phase
$\varphi$, the edge electrons also acquire a dynamical phase
$\lambda=2k_F\ell$ identical for both spin polarizations, where
$\ell$ is the distance travelled by the edge electrons from left
to right junction and $k_F$ is the edge state Fermi wave vector.
Details of the analysis are presented in Appendix A; here we
discuss only the main results. We obtain the effective $2 \times
2$ device scattering matrix $S$,
\begin{equation}\label{Seffective}
S(\phi,\theta)=\left(1-e^{i\lambda}\Phi r'_L(0)\Phi r_R(\theta)\right)^{-1}e^{i\lambda/2}\Phi,
\end{equation}
where the junction reflection matrices $r'_L(\theta_L)$ and
$r_R(\theta_R)$ depend on the angles $\theta_L,\theta_R$ of the
magnetization $\b{M}_{L,R}$ in the left and right leads. For
simplicity we consider $\theta_L=0$ and define
$\theta\equiv\theta_R$ (Fig. 1).

The two-terminal conductance $G$ of the device can be written as
\begin{equation}\label{landauer}
G=\frac{e^2}{h}\tr\rho_R S\rho_L S^\dag,
\end{equation}
using equation (\ref{trans}) of Appendix A. Here $\rho_L,\rho_R$
are $2\times 2$ effective spin density matrices for the FM leads,
and have the form
\begin{equation}\label{pLpR}
\rho_\alpha(\theta_\alpha)=\sh
T_\alpha(\theta_\alpha)\bigl(1+\b{P}_\alpha(\theta_\alpha)\cdot\bsigma\bigr),
\end{equation}
with $\alpha=L,R$, where $T_\alpha=\tr \rho_\alpha$ is the
transmission coefficient of the junction and $\b{P}_\alpha$ is a
polarization vector. For simplicity, we can assume the device to
have a $\pi$-rotation symmetry, which together with TR symmetry
restricts the generic form of the reflection matrices $r'_L$ and
$r_R$ in equation (\ref{Seffective}) to be
\begin{equation}\label{rLrR}
r_L'(\theta)=\left(
\begin{array}{cc}
\alpha_\theta & \beta_\theta \\
\gamma_\theta & \alpha_{\theta+\pi}
\end{array}
\right),
r_R(\theta)=\left(
\begin{array}{cc}
\alpha_{\theta+\pi} & \beta_\theta \\
\gamma_\theta & \alpha_\theta
\end{array}
\right).
\end{equation}
Physically, $\alpha_\theta$ is a non-spin-flip reflection
amplitude whereas $\beta_\theta,\gamma_\theta$ are spin-flip
reflection amplitudes, with $\beta_\theta$ corresponding to a
$\left|\d\rangle\right.\rightarrow\left|\u\rangle\right.$
reflection and $\gamma_\theta$ to a
$\left|\u\rangle\right.\rightarrow\left|\d\rangle\right.$
reflection. These amplitudes are generally different due to the
breaking of TR symmetry at the junctions by the nearby FM leads.

\section{Minimal model description}

These expressions being so far very general, to make further
progress it is useful to consider a simple continuum Hamiltonian
model for the FM/QSH junctions in which the reflection matrices
$r_L'$ and $r_R$ can be calculated explicitly. This model
satisfies the symmetries invoked earlier and will be seen to be a
good description of the realistic HgTe system in spite of its
simplicity. We model the FM leads as 1D spin-$\sh$ fermions with a
term which explicitly breaks the $SU(2)$ spin rotation symmetry
\cite{Slonczewski1989},
\[
H_\textrm{FM}=\int dx\,\Psi^\dag\left(-\frac{1}{2m}\frac{\partial^2}{\partial x^2}-\b{M}(\theta)\cdot\bsigma\right)\Psi,
\]
where $\b{M}(\theta)=M\hat{\b{n}}$, with
$\hat{\b{n}}=\hat{\b{x}}\cos\theta+\hat{\b{y}}\sin\theta$, is an
in-plane magnetization vector and $\Psi$ is a two-component spinor
$\Psi\equiv\left(\begin{array}{cc}\psi_\uparrow &
\psi_\downarrow\end{array}\right)^T$. In the absence of AB flux,
the QSH edge liquid consists of 1D massless helical fermions
\cite{wu2006,xu2006}. When the spins of the edge states are
polarized along the $z$ direction, the Hamiltonian is given by
\[
H_\textrm{QSH}=-i v\sum_{\alpha=t,b}\eta_\alpha\int dx\left(\psi_{\alpha\u}^\dag\partial_x\psi_{\alpha\u}-\psi_{\alpha\d}^\dag\partial_x\psi_{\alpha\d}\right),
\]
where $v$ is the edge state velocity and $\alpha=t,b$ refers to
the top and bottom edge, respectively, with $\eta_t=1$ and
$\eta_b=-1$.

\begin{figure}
\begin{center}
\includegraphics[width=3.5in]{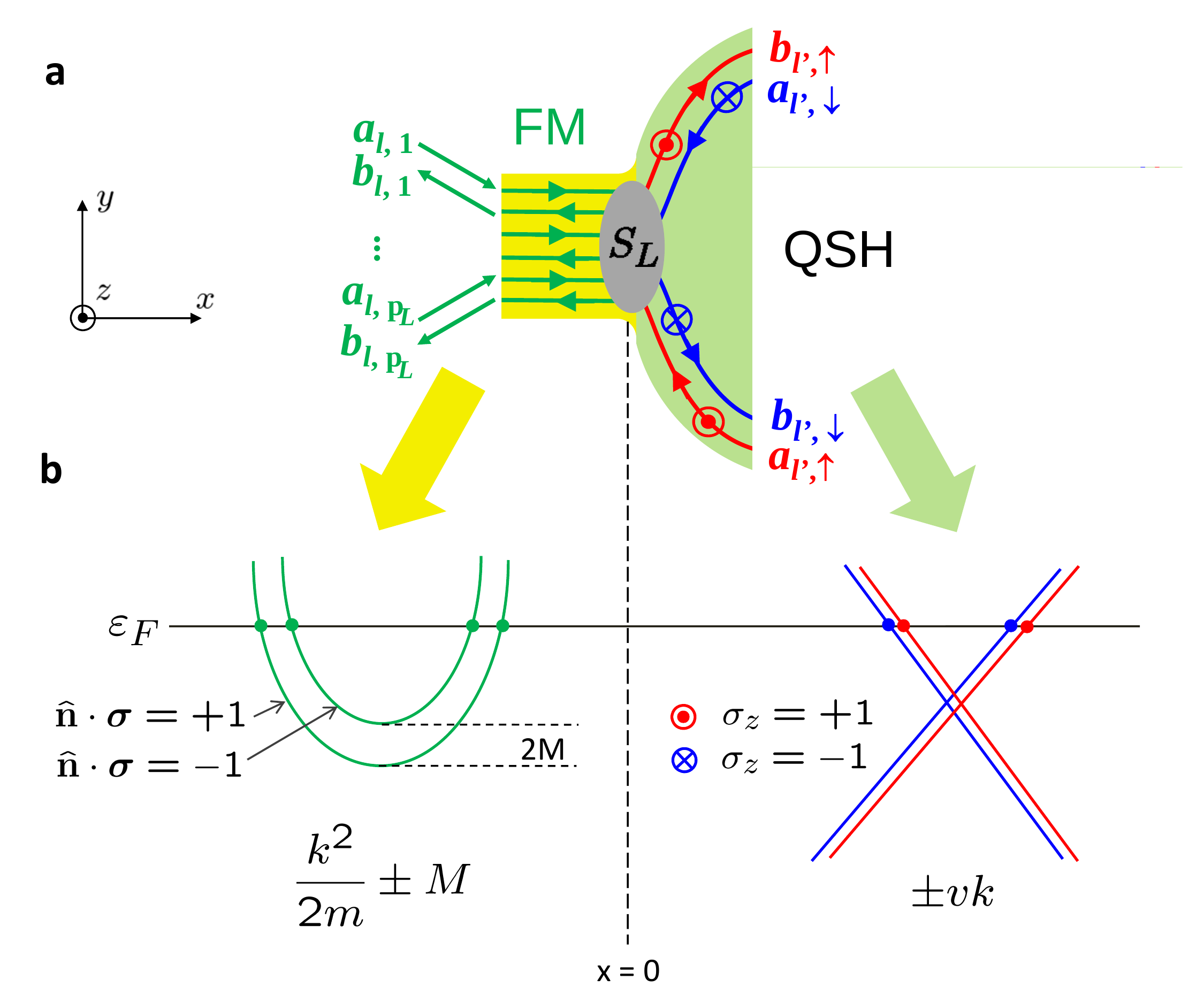}
\end{center}
\caption{{\bf Illustration of the minimal model describing a FM/QSH
junction.} {\bf a}, Schematic picture of the junction between the
left FM lead and the QSH insulator. Incoming channels
$a_{l,1},\ldots,a_{l,p_L}$ from the left lead scatter at the
junction into transmitted QSH edge channels $b_{l',\uparrow},
b_{l',\downarrow}$ and reflected lead channels
$b_{l,1},\ldots,b_{1,p_L}$. This scattering process is described by
a scattering matrix $S_L$. {\bf b}, Minimal model description of the
junction. The FM lead is described by 1D parabolic bands with a spin
splitting $2M$, while the QSH edge states are linearly dispersing and
TR invariant, with opposite spin states counter-propagating. %Band dispersion for simple 1D continuum model
%at the left junction: spin-split parabolic bands in FM leads (left
%side) and linearly dispersing in-gap helical edge states in the QSH
%insulator (right side). Both sides have the same number of
%propagating modes at the Fermi level $\EF>M$ where $2M=\Delta_s$ is
%the spin splitting in the FM leads.
} \label{fig:bands}
\end{figure}

In this simple model, the junction is described as a sharp
interface between the FM region and the QSH region, from which the
reflection matrix $r_L'$ in equation (\ref{Seffective}) and the
spin density matrix $\rho_L$ in equation (\ref{landauer}) can be
obtained. The calculation yields the reflection matrices precisely
in the form of equation (\ref{rLrR}) with $\alpha_\theta=a$ and
$\beta_\theta=\gamma_\theta^*=be^{-i\theta}$. In the limit of
small spin splitting $M/\EF\ll 1$ where $\varepsilon_F$ is the
Fermi energy in the leads (Fig. 2b), the reflection amplitudes $a$
and $b$ are given by
\begin{equation}\label{ab}%Joseph, I did not introduce eta, just to simplify the notation for the reader. Please doublecheck that the equations I modified are correct.
a\simeq\frac{v-v_F}{v+v_F}, \hspace{5mm}
b\simeq\frac{M}{2\varepsilon_F}\frac{v_F^3}{v(v+v_F)^2},
\end{equation}
where $v_F=\sqrt{2\varepsilon_F/m}$ is the Fermi velocity in the
FM leads. The off-diagonal spin-flip reflection amplitude $b$ is
proportional to the magnetization $M$ and along with its
accompanying scattering phase shift $e^{\pm i\theta}$ is an
explicit signature of TR symmetry breaking at the junction. The
diagonal non-spin-flip reflection amplitude $a$ does not break TR
symmetry and is the same as would be obtained in the scattering
from a nonmagnetic metal with $\b{M}=0$. The lead spin density
matrices $\rho_L,\rho_R$ can also be calculated explicitly and are
found to follow the form of equation (\ref{pLpR}) as expected from
the general $S$-matrix analysis. In the limit $M/\EF\ll 1$, we
obtain $T_L=T_R=8vv_F/(v+v_F)^2$ and
\begin{equation}\label{P}
\b{P}_L(\theta)=\b{P}_R(\theta)\equiv\b{P}(\theta)=-\frac{\b{M}(\theta)}{4\varepsilon_F}
\frac{v_F^2}{v(v+v_F)},
\end{equation}
i.e. the spin polarization vector is directly proportional to the
magnetization $\b{M}$.

\begin{figure}
\begin{center}
\includegraphics[width=3.5in]{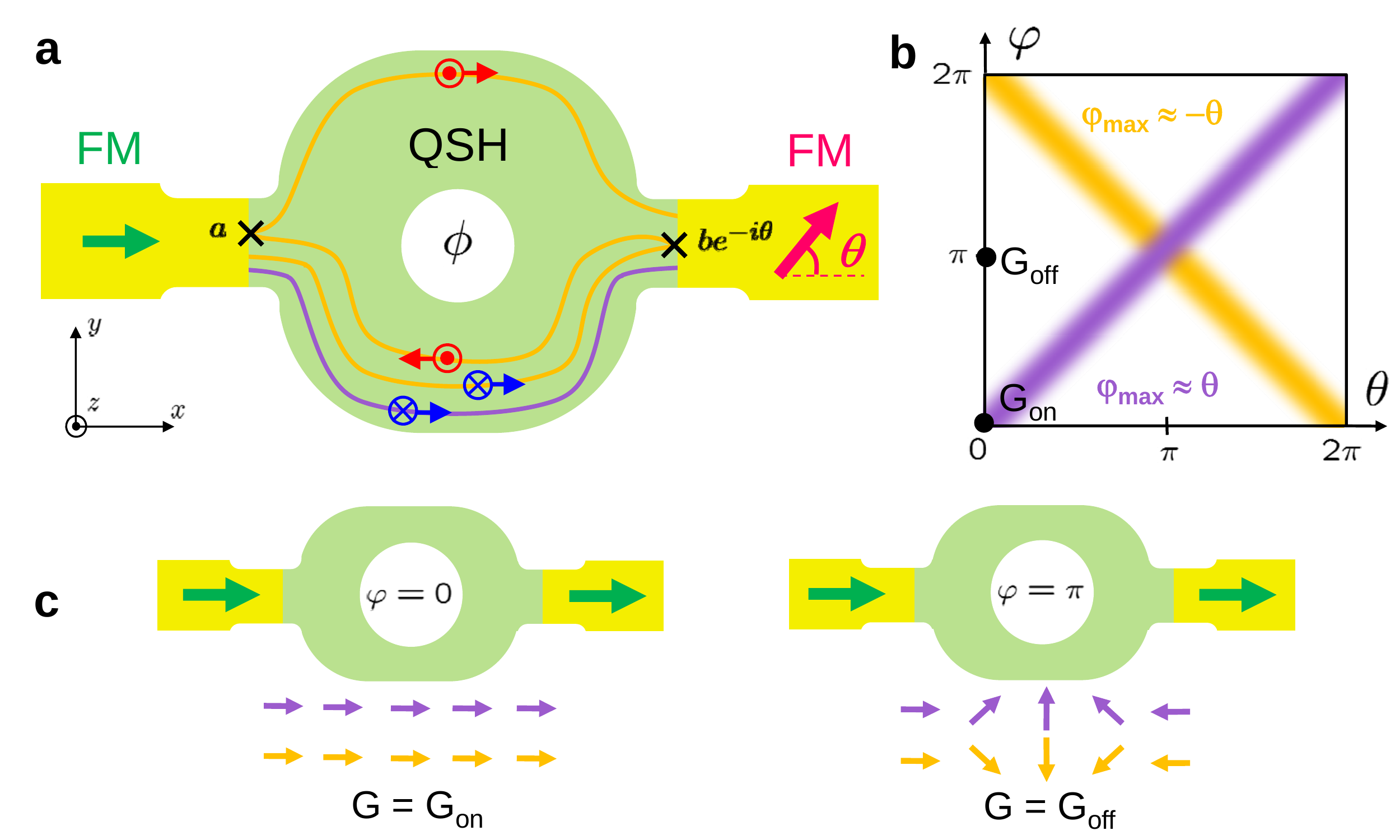}
\end{center}
\caption{{\bf Phenomenological analysis of the two-terminal
conductance} (top view of Fig. \ref{fig:2terminal}). {\bf a}, The two leading contributions to the spin AB
rotation. The purple path stands for the process with no spin flips,
which leads to a spin rotation of $\varphi\equiv2\pi\phi/\phi_0$.
The orange path stands for the process with spin-dependent
reflections, which leads to a spin rotation of $-\varphi$. {\bf b},
Schematic intensity map of the two-terminal conductance
$G(\varphi,\theta)$. The conductance reaches its maximum along the
lines $\varphi=\theta$ (purple) and $\varphi=-\theta$ (orange),
which are contributed by the purple and orange paths in panel {\bf
a}, respectively. {\bf c}, The on and off states of the topological
spin transistor are defined for $\theta=0$ by $\varphi=0$ and $\varphi=\pi$,
respectively, as also indicated in panel {\bf b}.
%{\bf a}, Generic intensity map of the two-terminal conductance:
%processes in which electrons traverse the device without suffering
%spin-flip reflections at the junctions (purple line) yield a large
%conductance along the line $\varphi=\theta$, while processes
%involving spin-flip reflections (orange line) yield a large
%conductance along the line $\varphi=-\theta$. {\bf b}, Process with
%no spin flips (purple) contributing to the $\varphi=\theta$
%intensity and a process involving a spin-flip reflection (orange)
%contributing to the $\varphi=-\theta$ intensity. {\bf c}, Depiction
%of the on and off states of the topological spin
%transistor.
}\label{fig:crossing}
\end{figure}

From the results obtained above, we can readily evaluate the
conductance $G$, which has the following expression in the limit
$M/\EF, P\equiv|{\bf P}(\theta)|\ll 1$ and $\lambda=0$: %Joseph, do we need to mention the effect of dynamical phase somewhere after the following equation?
\begin{eqnarray}\label{G1D}
\lefteqn{G(\varphi,\theta;\lambda=0)=\frac{e^2}{h}\frac{T_LT_R/2}
{1-2a^2\cos\varphi+a^4}}\nonumber\\
&&\times\biggl[1
+\frac{\cos(\theta-\varphi)+(1-t^2)^2\cos(\theta+\varphi)+C(\varphi,\theta)}
{1-2a^2\cos\varphi+a^4}P^2\nonumber\\
&&+\c{O}(P^4)\biggr],
\end{eqnarray}
where $t=1-a$ and
$C(\varphi,\theta)\equiv\gamma\cos\varphi+\delta\cos\theta$ with
$\gamma,\delta$ some constants depending only on $a$. The effect
of a finite $\lambda$ will be addressed in the next section, where
we study numerically a more realistic model of the QSH state in
HgTe QWs. Physically, $a$ and $t$ can be interpreted as reflection
and transmission coefficients for the $S_z$ spin current. The
generic behavior of equation (\ref{G1D}) is illustrated in Fig.~3.
The term $C(\varphi,\theta)$ is an uninteresting background term
which manifests no correlation between AB phase $\varphi$ and
rotation angle of the electron spin $\theta$. The term
$\propto\cos(\theta-\varphi)$ corresponds to a rotation of the
electron spin by $\varphi$, and the term
$\propto\cos(\theta+\varphi)$ corresponds to a rotation by
$-\varphi$. The conductance is thus maximal for
$\varphi_\mathrm{max}=\pm\theta$ (Fig.~3 b), manifesting the
desired flux-induced spin rotation effect. Physically, the
$\varphi_\mathrm{max}=\theta$ term corresponds to a process in
which electrons traverse the device without undergoing spin flips
(Fig.~3 a, purple trajectory), while the
$\varphi_\mathrm{max}=-\theta$ term corresponds to a process
involving at least one TR breaking spin-flip reflection (Fig.~3 a,
orange trajectory). As can be seen from equation (\ref{G1D}), the
relative intensity of the two contributions to the conductance is
$I_{-\theta}/I_\theta=(1-t^2)^2$ which can be close to unity for
strongly reflecting junctions $t\ll 1$. As both contributions are
minimal for $\varphi=\pi$ at $\theta=0$, one can consider
$\varphi=\pi$, $\theta=0$ as the `off' state of a spin transistor
(Fig.~3 c, right) where the rotation of the spin is provided by a
purely quantum mechanical Berry phase effect. This is in contrast
with the famous Datta-Das spin transistor \cite{Datta1990} where
the rotation of the spin is achieved through the classical
spin-orbit force. The `on' state corresponds to the absence of
spin rotation for $\varphi=0$ (Fig.~3 c, left).

\section{Experimental realization in $\mathrm{HgTe}$ quantum wells}

We now show that this proposal can in principle be realized
experimentally in HgTe QWs. We model the device of Fig.~1 as a
rectangular QSH region threaded by a magnetic AB flux through a
single plaquette in the center, and connected to semi-infinite
metallic leads on both sides by rectangular QSH constrictions
modeling quantum point contacts (QPC) (Fig.~4a). The QSH region is
described by an effective $4\times 4$ tight-binding Hamiltonian
\cite{bernevig2006d,Konig2008} with the chemical potential in the
bulk gap, while the metallic leads are described by the same model
with the chemical potential in the conduction band. The detailed
form of the model is given in Appendix C. The injection of
spin-polarized carriers by the FM layers of Fig.~1 is mimicked by
the inclusion of an effective Zeeman term in the Hamiltonian of
the semi-infinite leads. We calculate numerically the two-terminal
conductance through the device of Fig.~4a for a QW thickness
$d=80$~\AA. We use the standard lattice Green function
Landauer-B\"{u}ttiker approach \cite{FerryGoodnick} in which the
conductance is obtained from the Green function of the whole
device, the latter being calculated recursively \cite{Wu1994}.

\begin{figure}
\begin{center}
\includegraphics[width=3.5in]{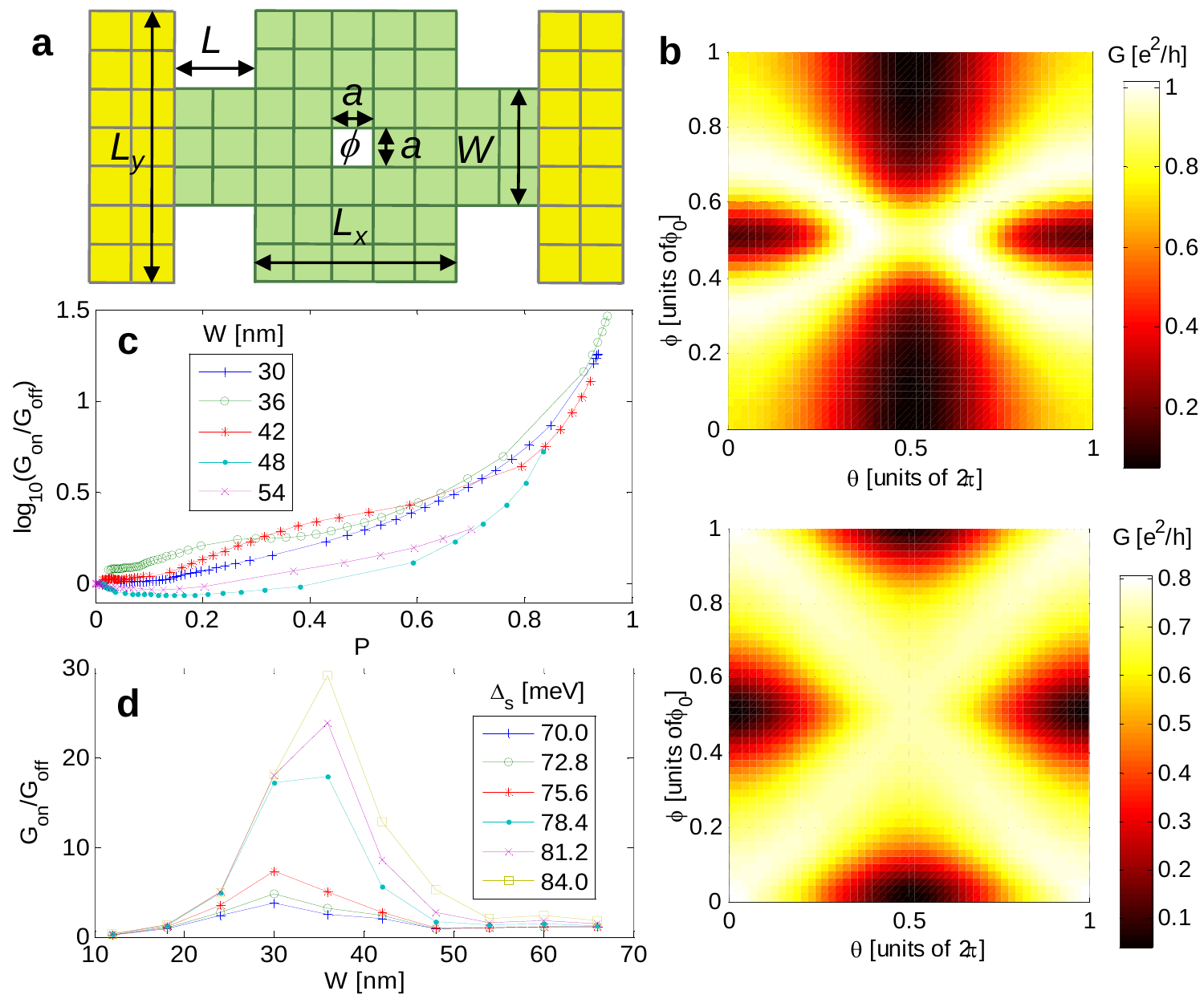}
\end{center}
\caption{{\bf Numerical study of the spin AB effect in HgTe QWs.}
{\bf a}, Device geometry used for the numerical two-terminal
conductance calculation: $a=30$ \AA~is the lattice constant of the
tight-binding model, $L=18\,\mathrm{nm}$,
$L_x=\ell=240\,\mathrm{nm}$, $L_y=120\,\mathrm{nm}$, $\phi$ is the
AB flux, and $W$ is the QPC width. {\bf b}, Intensity map of the
conductance $G(\phi,\theta)$ for fixed chemical potential
$\mu=0.06\,\mathrm{eV}$ (top panel) and averaged chemical potential
over energy range $\Delta\mu=5\,\mathrm{ meV}$ corresponding to an
average over $\sim 2\pi$ dynamical phase (bottom panel). These two
situations correspond to low and high temperature, respectively (see
text). {\bf c}, Logarithmic plot of on/off ratio
$G_\mathrm{on}/G_\mathrm{off}$ of topological spin transistor as a
function of spin polarization $P$ of injected carriers for fixed
chemical potential $\mu=0.06\,\mathrm{eV}$ and different values of
the QPC width $W$. {\bf d}, Plot of on/off ratio as a function of
QPC width $W$ for fixed chemical potential $\mu=0.06\,\mathrm{eV}$
and different values of the spin splitting $\Delta_s$ in the bulk
leads. }\label{fig:numerics}
\end{figure}

The results of the numerical calculation are plotted in Fig.~4 b, c,
d. In the absence of phase-breaking scattering processes, one
distinguishes two temperatures regimes $T\ll T_\ell$ and $T\gg
T_\ell$ separated by a crossover temperature $T_\ell=\pi\hbar
v/k_B\ell$ with $v$ the edge state velocity, defined as the
temperature for which a thermal spread $\Delta\mu\sim k_BT$ in the
energy distribution of injected electrons corresponds to a spread in
the distribution of dynamical phases $\lambda=2k_F\ell$ of
$\Delta\lambda\sim 2\pi$. In the low temperature regime $T\ll
T_\ell$, $\Delta\lambda\ll 2\pi$ and the dynamical phase is
essentially fixed such that $G(T\ll T_\ell)\simeq G(T=0)$. In this
regime, $G(T=0,\mu)$ is approximately periodic in $\mu$ for $\mu$
within the bulk gap, with period $\Delta\mu\sim k_BT_\ell$. A
crossing pattern (Fig.~4b, top) occurs periodically and can be
obtained by tuning the chemical potential. It corresponds to the
flux-induced spin rotation effect (Fig.~3). In the high temperature
regime $T\gg T_\ell$, one could expect that the crossing pattern,
and thus the spin rotation effect, would be washed out by thermal
self-averaging of the dynamical phase. Surprisingly, the pattern
remains (Fig.~4b, bottom), and actually acquires a more symmetric
structure through the self-averaging procedure. In both temperature
regimes, the conductance pattern agrees qualitatively with the
result of the simple 1D Hamiltonian model (Fig.~3 b).

In the presence of phase-breaking scattering processes, the upper
bound for the system size $\ell$ is given by the phase coherence
length $\ell_\varphi$, which defines a minimum crossover
temperature $T_\ell^\mathrm{min}=\pi\hbar v/k_B\ell_\varphi$. The
self-averaging regime can in principle be reached for $T\gg
T_\ell$; however, at too high temperatures one expects
phase-breaking processes to reduce $\ell_\varphi$ below the system
size and the QSH state can be destroyed. In HgTe QWs one estimates
\cite{koenig2007b} $\ell_\varphi\sim 1\,\mu$m, such that for a
typical edge state velocity $\hbar v\sim 3.5$~eV$\cdot$\AA~one
obtains a minimum crossover temperature $T_\ell^\mathrm{min}\sim
13$~K. Therefore, for a device of size $\ell\lesssim 1\,\mu$m one
should be able to tune the spin rotation crossing pattern with the
chemical potential for measurement temperatures $T\ll 13$ K. At
high temperatures $T\gg 13$ K, the QSH state is presumably
destroyed due to increased amounts of inelastic phase-breaking
processes and the self-averaging regime cannot be reached.
However, in type-II QWs \cite{Liu2008} the edge state velocity is
about one order of magnitude smaller, hence $T_\ell\sim 1$ K and
it might be possible to reach the self-averaging high temperature
regime $T\gtrsim 1$ K without destroying the phase coherence of
the sample.

In our calculations, for simplicity we have assumed that electrons
on both the top and bottom edges acquire the same dynamical phase
$\lambda$. In a real system, the two arms of the ring are not
perfectly symmetric and the electrons propagating on different
arms can certainly acquire different dynamical phases
$\lambda_\mathrm{bottom}\neq\lambda_\mathrm{top}$. However, the
dynamical phase difference
$\delta\equiv\lambda_\mathrm{bottom}-\lambda_\mathrm{top}$ only
leads to an additional flux-independent rotation of the spin of
the outgoing electrons, which leads to a shift of the conductance
pattern in the angle $\theta$ by an amount $\delta$ (see equation
(\ref{Gshift}) of Appendix A). Thus the transistor remains
effective if one uses $\theta=\delta$ instead of $\theta=0$ in the
right FM lead. If one prefers to use $\theta=0$, one can cancel
out the phase asymmetry by patterning an electrostatic gate on top
of one given arm. By tuning the potential of this gate, one can
adjust the Fermi wave vector locally and introduce a dynamical
phase offset which cancels out the phase asymmetry $\delta$.

In Fig. 4c,d we plot the on/off ratio $G_\mathrm{on}/G_\mathrm{off}$
of the topological spin transistor, which can be taken as the figure
of merit of the device. We define $G_\mathrm{on}\equiv
G(\phi=0,\theta=0)$ and $G_\mathrm{off}\equiv
G(\phi=\frac{1}{2}\phi_0,\theta=0)$ (see Fig.~3 c). We use two
parameters, the junction spin polarization $P$ and the bulk spin
splitting $\Delta_s$ to quantify the degree of spin polarization of
the injected carriers. An actual experimental implementation of the
transistor concept described here will require optimization of these
or similar parameters. The junction spin polarization $P$ is
obtained for a given junction geometry, i.e. a given choice of QPC
width $W$ and length $L$ (Fig.~4a), by calculating the transfer
matrix \cite{Sanvito1999} of the junction directly from the TB model
and using equation (\ref{pLpR}) with $P\equiv|\b{P}|$. The spin
splitting $\Delta_s$ is obtained from the continuum
$\b{k}\cdot\b{p}$ HgTe QW Hamiltonian mentioned earlier, and is
defined as the energy difference between `spin up' ($E1+$) and `spin
down' ($E1-$) energy levels \cite{Konig2008} at the $\Gamma$ point.
The on/off ratio increases rapidly for a polarization $P$ of order
unity (Fig.~4c). It is reasonable to expect that optimized junction
designs, better that the simplistic proof-of-concept geometry used
here, would yield even higher on/off ratios. There is also an
optimal width $W_\mathrm{opt}\simeq~0.29L_y$ for the junction QPC
(Fig.~4d). For $W< W_\mathrm{opt}$, interedge tunneling
\cite{Zhou2008} strongly backscatters the incoming electrons and
reduces $G_\mathrm{on}$, which suppresses the on/off ratio. For $W>
W_\mathrm{opt}$, the edge states on opposite edges are too far apart
to recombine coherently and to produce the desired spin rotation
effect, which increases $G_\mathrm{off}$ and also suppresses the
on/off ratio.

\section{Conclusion and outlook}

In this work, we have shown the possibility of using a
topologically nontrivial state of matter, the QSH insulator state,
to manipulate the spin of the electron by purely nonlocal, quantum
mechanical means, without recourse to local interactions with
classical electromagnetic fields. This spin AB effect, which is a
spin analog of the usual charge AB effect, relies on the helical
and topological nature of the QSH edge states which is peculiar to
that state of matter, combined with a Berry phase effect. In
addition, we have shown that the spin AB effect can be used to
design a new kind of spin transistor which is fundamentally
different from the previous proposals, in that there is no
classical force or torque acting on the spin of the electron.
Furthermore, edge transport in the QSH regime being
dissipationless \cite{koenig2007b,Roth2009,Buttiker2009}, the
proposed topological spin transistor would have the advantage of a
lower power consumption in comparison to previous proposals for
spin transistors. More generally, such a quantum manipulation of
the electron spin, if observed, could open new directions in
spintronics research and applications, and would at the same time
demonstrate the practical usefulness of topological states of
quantum matter.

We are especially grateful to S.-C. Zhang for many illuminating
discussions and collaborations at the early stage of this project.
We also thank C.-C. Chen, B. Huard, T. L. Hughes, M. K\"{o}nig,
C.-X. Liu, Y. Oreg, S. Raghu and H. Yao for insightful
discussions. J.M. is supported by the National Science and
Engineering Research Council (NSERC) of Canada, the Fonds
qu\'{e}b\'{e}cois de la recherche sur la nature et les
technologies (FQRNT), and the Stanford Graduate Program (SGF).
E.-A.K. is supported in part by the Nanoscale Science and
Engineering Initiative of the National Science Foundation under
NSF Award \#EEC-0646547. X.-L.Q. is supported by the Department of
Energy, Office of Basic Energy Sciences, Division of Materials
Sciences and Engineering, under contract DE-AC02-76SF00515.

{\em Note added}.---After the completion of this work, we became
aware of a recent preprint by Usaj\cite{usaj2009} which discusses
a similar effect in a different physical system.

%\section{Methods}\label{methods}
\appendix

\section{$S$-matrix analysis}

We wish to obtain an expression for the $S$-matrix $\bigS$
relating outgoing $b$ to incoming $a$ current amplitudes,
\begin{equation}\label{Smatrix}
\left(
\begin{array}{c}
b_l\\
b_r
\end{array}
\right)=
\bigS
\left(
\begin{array}{c}
a_l\\
a_r
\end{array}
\right)\text{ with }
\bigS=\left(
\begin{array}{cc}
r & t'\\
t & r'
\end{array}
\right),
\end{equation}
where $a_l$ and $b_l$ ($a_r$ and $b_r$) are $p_L\times 1$
($p_R\times 1$) column vectors of the current amplitudes outside
the QSH region in the left (right) lead (see Fig. 1), and $p_L$
($p_R$) is the number of propagating channels at the Fermi energy
in the left (right) lead. The matrix $\bigS$ therefore has
dimensions $(p_L+p_R)\times(p_L+p_R)$ and the submatrices $r,r'$
and $t,t'$ are reflection and transmission matrices, respectively.
The two-terminal conductance $G$ from left to right is given by
the Landauer formula \cite{FerryGoodnick} $G=\frac{e^2}{h}\tr
tt^\dag$. We assume that phase coherence is preserved throughout
the sample so that $\bigS$ can be obtained by combining
$S$-matrices for different portions of the device coherently
\cite{FerryGoodnick}. We define the $(p_{L,R}+2)\times(p_{L,R}+2)$
scattering matrices $S_L,S_R$ for the left ($L$) and right ($R$)
FM/QSH junctions (e.g. see Fig. 2a for the left junction),
\begin{equation}\label{SLSR}
\left(
\begin{array}{c}
b_l\\
b_{l'}
\end{array}
\right)=
S_L\left(
\begin{array}{c}
a_l\\
a_{l'}
\end{array}
\right),
\hspace{5mm}
\left(
\begin{array}{c}
b_{r'}\\
b_r
\end{array}
\right)=
S_R\left(
\begin{array}{c}
a_{r'}\\
a_r
\end{array}
\right),
\end{equation}
where $l'$ ($r'$) is the QSH region immediately to the right
(left) of the left (right) junction, such that $a_{l'},a_{r'}$ and
$b_{l'},b_{r'}$ are the 2-component spinors of edge state current
amplitudes. They are related through the geometric AB phase
$\varphi$ (different for each spin polarization) and the dynamical
phase $\lambda=2k_F\ell$ (identical for both spin polarizations)
where $\ell$ is the distance travelled by the edge electrons from
left to right junction and $k_F$ is the edge state Fermi
wavevector,
\begin{equation}\label{relations}
\left\{
\begin{array}{c}
a_{r'\u,\d}\\
a_{l'\u,\d}
\end{array}
\right\}=e^{i\lambda/2}e^{\mp i\varphi/2}
\left\{
\begin{array}{c}
b_{l'\u,\d}\\
b_{r'\u,\d}
\end{array}
\right\},
\end{equation}
where the upper sign for $\varphi$ corresponds to spin up. Using
equations (\ref{SLSR}) and (\ref{relations}), we can write
\begin{equation}\label{SR_Phi}
\left(
\begin{array}{c}
e^{-i\lambda/2}\Phi^\dag a_{l'}\\
b_r
\end{array}
\right)=
S_R\left(
\begin{array}{c}
e^{i\lambda/2}\Phi b_{l'}\\
a_r
\end{array}
\right),
\end{equation}
where we define $\Phi\equiv e^{-i\varphi\sigma_z/2}$. Using the
first equality in equation (\ref{SLSR}) together with equation
(\ref{SR_Phi}), we can eliminate the intermediate amplitudes
$a_{l'},b_{l'}$ and obtain relations between the left lead
amplitudes $a_l,b_l$ and the right lead amplitudes $a_r,b_r$,
which gives us $\c{S}$ (equation (\ref{Smatrix})). The $2\times 2$
transmission matrix $t$, i.e. the lower left block of $\c{S}$, is
then obtained in the form
\begin{equation}\label{trans}
t=t_R S t_L,
\end{equation}
where $t_L$ and $t_R$ are the $2\times p_L$ and $p_R\times 2$
transmission matrices for the left and right junctions,
respectively (i.e. the lower left blocks of $S_L,S_R$ following
the notation of equation (\ref{Smatrix})), and $S$ is a $2\times
2$ matrix defined in equation (\ref{Seffective}). The effective
spin density matrices $\rho_L,\rho_R$ of the FM leads used in
equation (\ref{landauer}) are defined as $\rho_L=t_Lt_L^\dag$ and
$\rho_R=t_R^\dag t_R$.

If the arms of the ring are asymmetric, the dynamical phase
$\lambda$ is generally different for each arm and we have
$\lambda_\mathrm{bottom}-\lambda_\mathrm{top}\equiv\delta\neq 0$.
In this case, one can show that equation (\ref{landauer}) still
holds, but with the substitutions
\begin{eqnarray*}
\rho_L(\theta_L)&\rightarrow& R_\delta\rho_L(\theta_L) R_\delta^{-1}
=\rho_L(\theta_L+\delta),\\
r'_L(\theta_L)&\rightarrow& R_\delta r'_L(\theta_L)R_\delta^{-1}
=r'_L(\theta_L+\delta),
\end{eqnarray*}
where $R_\delta\equiv e^{-i\sigma_z\delta/2}$ rotates the spin
about the $z$ axis by an angle $\delta$. In other words, a phase
asymmetry is equivalent to a rigid flux-independent rotation of
the electron spin, and simply shifts the conductance pattern by a
constant angle $\delta$:
\begin{eqnarray}\label{Gshift}
G(\phi,\theta\equiv\theta_R-\theta_L)&\rightarrow&
G(\phi,\theta_R-(\theta_L+\delta))\nonumber\\
&=&G(\phi,\theta-\delta).
\end{eqnarray}

\section{Scattering at the junction} In order to solve the 1D
scattering problem at the FM/QSH interface, we first observe that
the number of degrees of freedom is equal on either side of the
junction. If the Fermi level $\EF$ is chosen such that both spin
subbands in the FM leads are occupied, there are four propagating
modes on each side of the junction (two spins and two chiralities).
The QSH spin states $\phi^{\textrm{QSH}(\pm)}$ are $\sigma_z$
eigenstates while the FM spin states
$\phi^{\textrm{FM}(\pm)}(\theta)$ are eigenstates of
$\hat{\b{n}}\cdot\bsigma$ and depend explicitly on $\theta$. The
Schr\"odinger equation for the junction is then solved by the
following scattering ansatz,
\[
\psi_\sigma^{(+)}(x)=\left\{
\begin{array}{cc}
\frac{\phi_\sigma^{<(+)}}{\sqrt{v_\sigma^<}}e^{ik_\sigma^<x}+\sum_{\sigma'}r_{\sigma'\sigma}\frac{\phi_{\sigma'}^{<(-)}}{\sqrt{v_{\sigma'}^<}}e^{-ik_{\sigma'}^<x}, & x<0, \\
\sum_{\sigma'}t_{\sigma'\sigma}\frac{\phi_{\sigma'}^{>(+)}}{\sqrt{v_{\sigma'}^>}}e^{ik_{\sigma'}^>x}, & x>0,
\end{array}
\right.
\]
for a right-moving scattering state, and with similar expressions
for a left-moving scattering state $\psi_\sigma^{(-)}$. Spin is
denoted by $\sigma$, chirality by $\pm$ and side of the junction
by $<,>$. The propagating modes are explicitly normalized to unit
flux such that $r_{\sigma'\sigma}$ and $t_{\sigma'\sigma}$ are the
desired reflection and transmission matrices. Requiring the
continuity of $\psi_\sigma^{(\pm)}$ and
$\hat{v}_x\psi_\sigma^{(\pm)}$ at the interface $x=0$ (with
$\hat{v}_x\equiv\partial H/\partial k_x$ the velocity operator),
we obtain a system of linear equations for the sixteen matrix
elements $r_L,t_L,r'_L,t'_L$ constituting $S_L$. As illustrated in
Fig. \ref{fig:2terminal}, the magnetization angle is set to zero
in the left lead and to $\theta$ in the right lead and we obtain
$r_L'(0)$ and $r_R(\theta)$ in equation (\ref{Seffective}).

\section{Tight-binding model}

The effective tight-binding model describing HgTe QWs is defined
as \cite{bernevig2006d,Konig2008}
\begin{equation}\label{TB}
\c{H}=\sum_i c^\dag_i V_i c_i+\sum_{ij}\Bigl(c_i^\dag T_{ij}
e^{iA_{ij}}c_j+\mathrm{h.c.}\Bigr),
\end{equation}
where
$T_{ij}=T_{\hat{x}}\delta_{j,i+\hat{x}}+T_{\hat{y}}\delta_{j,i+\hat{y}}$
is the nearest-neighbor hopping matrix, $A_{ij}=\frac{e}{\hbar
c}\int_i^jd\b{r}\cdot\b{A}$ is the Peierls phase with $\b{A}$ the
electromagnetic vector potential, and $V_i$, $T_{\hat{x}}$ and
$T_{\hat{y}}$ are $4\times 4$ matrices containing the
$\b{k}\cdot\b{p}$ parameters and the effective Zeeman term. The
$4\times 4$ matrices $T_{\hat{x}}$, $T_{\hat{y}}$ and $V_i$ used
in the tight-binding Hamiltonian (\ref{TB}) are given by
\begin{eqnarray}\label{TxTy}
T_{\hat{x}}&=&\left(
\begin{array}{cccc}
D_+ & -\frac{iA}{2} & -\frac{i\Delta_e}{2} & 0 \\
-\frac{iA}{2} & D_- & 0 & -\frac{i\Delta_h}{2} \\
-\frac{i\Delta_e}{2} & 0 & D_+ & \frac{iA}{2} \\
0 & -\frac{i\Delta_h}{2} & \frac{iA}{2} & D_-
\end{array}
\right),\nonumber\\
T_{\hat{y}}&=&\left(
\begin{array}{cccc}
D_+ & \frac{A}{2} & \frac{\Delta_e}{2} & 0 \\
-\frac{A}{2} & D_- & 0 & -\frac{\Delta_h}{2} \\
-\frac{\Delta_e}{2} & 0 & D_+ & \frac{A}{2} \\
0 & \frac{\Delta_h}{2} & -\frac{A}{2} & D_-
\end{array}
\right),
\end{eqnarray}
and
\begin{eqnarray}\label{Vi}
V_i&=&(C-4D-\varepsilon_F+E_g(i))\mathds{1}_{4\times 4}\nonumber\\
&&+(M-4B)
\mathds{1}_{2\times 2}\otimes\sigma_z+
H_{Z\parallel}^\mathrm{eff}+H_{Z\perp}^\mathrm{eff},
\end{eqnarray}
where $D_\pm\equiv D\pm B$ and $A,B,C,D,M,\Delta_e,\Delta_h$ are
$\b{k}\cdot\b{p}$ parameters \cite{Konig2008}, and
$\mathds{1}_{n\times n}$ denotes the $n\times n$ unit matrix. The
Fermi energy $\varepsilon_F$ is uniform throughout the device. The
gate potential $E_g(i)$ is different in the QSH and lead regions
(Fig. 4a), and is used to tune the central region into the QSH
insulating regime. The in-plane $H_{Z\parallel}^\mathrm{eff}$ and
out-of-plane $H_{Z\perp}^\mathrm{eff}$ effective Zeeman terms,
which are used to mimick the injection of spin-polarized carriers
from a FM layer (Fig. 1), are given by \cite{Konig2008}
\begin{eqnarray}\label{HZ}
H_{Z\parallel}^\mathrm{eff}&=&g_\parallel \mu_B\left(
\begin{array}{cccc}
0 & 0 & B_-^\mathrm{eff} & 0 \\
0 & 0 & 0 & 0 \\
B_+^\mathrm{eff} & 0 & 0 & 0 \\
0 & 0 & 0 & 0
\end{array}
\right),\nonumber\\
H_{Z\perp}^\mathrm{eff}&=&\mu_B B_z^\mathrm{eff}\left(
\begin{array}{cccc}
g_{E\perp} & 0 & 0 & 0 \\
0 & g_{H\perp} & 0 & 0  \\
0 & 0 & -g_{E\perp} & 0 \\
0 & 0 & 0 & -g_{H\perp}
\end{array}
\right),
\end{eqnarray}
where $B_\pm^\mathrm{eff}=B_x^\mathrm{eff}\pm iB_y^\mathrm{eff}$,
$\b{B}^\mathrm{eff}=(B_x^\mathrm{eff},B_y^\mathrm{eff},B_z^\mathrm{eff})$
is some effective magnetic field whose role is to induce a spin
polarization in the leads, $\mu_B$ is the Bohr magneton, and
$g_\parallel$ and $g_{E\perp},g_{H\perp}$ are the in-plane and
out-of-plane $g$-factors, respectively.

\bibliography{dattadas}

\begin{thebibliography}{23}
\expandafter\ifx\csname natexlab\endcsname\relax\def\natexlab#1{#1}\fi
\expandafter\ifx\csname bibnamefont\endcsname\relax
  \def\bibnamefont#1{#1}\fi
\expandafter\ifx\csname bibfnamefont\endcsname\relax
  \def\bibfnamefont#1{#1}\fi
\expandafter\ifx\csname citenamefont\endcsname\relax
  \def\citenamefont#1{#1}\fi
\expandafter\ifx\csname url\endcsname\relax
  \def\url#1{\texttt{#1}}\fi
\expandafter\ifx\csname urlprefix\endcsname\relax\def\urlprefix{URL }\fi
\providecommand{\bibinfo}[2]{#2}
\providecommand{\eprint}[2][]{\url{#2}}

\bibitem[{\citenamefont{\v{Z}uti\'{c} et~al.}(2004)\citenamefont{\v{Z}uti\'{c},
  Fabian, and {Das Sarma}}}]{zutic2004}
\bibinfo{author}{\bibfnamefont{I.}~\bibnamefont{\v{Z}uti\'{c}}},
  \bibinfo{author}{\bibfnamefont{J.}~\bibnamefont{Fabian}}, \bibnamefont{and}
  \bibinfo{author}{\bibfnamefont{S.}~\bibnamefont{{Das Sarma}}},
  \bibinfo{journal}{Rev. Mod. Phys.} \textbf{\bibinfo{volume}{76}},
  \bibinfo{pages}{323} (\bibinfo{year}{2004}).

\bibitem[{\citenamefont{Aharonov and Bohm}(1959)}]{aharonov1959}
\bibinfo{author}{\bibfnamefont{Y.}~\bibnamefont{Aharonov}} \bibnamefont{and}
  \bibinfo{author}{\bibfnamefont{D.}~\bibnamefont{Bohm}},
  \bibinfo{journal}{Phys. Rev.} \textbf{\bibinfo{volume}{115}},
  \bibinfo{pages}{485} (\bibinfo{year}{1959}).

\bibitem[{\citenamefont{Kane and Mele}(2005{\natexlab{a}})}]{kane2005}
\bibinfo{author}{\bibfnamefont{C.~L.} \bibnamefont{Kane}} \bibnamefont{and}
  \bibinfo{author}{\bibfnamefont{E.~J.} \bibnamefont{Mele}},
  \bibinfo{journal}{Phys. Rev. Lett.} \textbf{\bibinfo{volume}{95}},
  \bibinfo{pages}{226801} (\bibinfo{year}{2005}{\natexlab{a}}).

\bibitem[{\citenamefont{Bernevig and Zhang}(2006)}]{bernevig2006a}
\bibinfo{author}{\bibfnamefont{B.~A.} \bibnamefont{Bernevig}} \bibnamefont{and}
  \bibinfo{author}{\bibfnamefont{S.~C.} \bibnamefont{Zhang}},
  \bibinfo{journal}{Phys. Rev. Lett.} \textbf{\bibinfo{volume}{96}},
  \bibinfo{pages}{106802} (\bibinfo{year}{2006}).

\bibitem[{\citenamefont{Murakami}(2006)}]{Murakami2006}
\bibinfo{author}{\bibfnamefont{S.}~\bibnamefont{Murakami}},
  \bibinfo{journal}{Phys. Rev. Lett.} \textbf{\bibinfo{volume}{97}},
  \bibinfo{pages}{236805} (\bibinfo{year}{2006}).

\bibitem[{\citenamefont{Bernevig et~al.}(2006)\citenamefont{Bernevig, Hughes,
  and Zhang}}]{bernevig2006d}
\bibinfo{author}{\bibfnamefont{B.~A.} \bibnamefont{Bernevig}},
  \bibinfo{author}{\bibfnamefont{T.~L.} \bibnamefont{Hughes}},
  \bibnamefont{and} \bibinfo{author}{\bibfnamefont{S.~C.} \bibnamefont{Zhang}},
  \bibinfo{journal}{Science} \textbf{\bibinfo{volume}{314}},
  \bibinfo{pages}{1757} (\bibinfo{year}{2006}).

\bibitem[{\citenamefont{Liu et~al.}(2008)\citenamefont{Liu, Hughes, Qi, Wang,
  and Zhang}}]{Liu2008}
\bibinfo{author}{\bibfnamefont{C.}~\bibnamefont{Liu}},
  \bibinfo{author}{\bibfnamefont{T.~L.} \bibnamefont{Hughes}},
  \bibinfo{author}{\bibfnamefont{X.-L.} \bibnamefont{Qi}},
  \bibinfo{author}{\bibfnamefont{K.}~\bibnamefont{Wang}}, \bibnamefont{and}
  \bibinfo{author}{\bibfnamefont{S.-C.} \bibnamefont{Zhang}},
  \bibinfo{journal}{Phys. Rev. Lett.} \textbf{\bibinfo{volume}{100}},
  \bibinfo{pages}{236601} (\bibinfo{year}{2008}).

\bibitem[{\citenamefont{Shitade et~al.}(2009)\citenamefont{Shitade, Katsura,
  {Kune\v{s}}, Qi, Zhang, and Nagaosa}}]{Shitade2009}
\bibinfo{author}{\bibfnamefont{A.}~\bibnamefont{Shitade}},
  \bibinfo{author}{\bibfnamefont{H.}~\bibnamefont{Katsura}},
  \bibinfo{author}{\bibfnamefont{J.}~\bibnamefont{{Kune\v{s}}}},
  \bibinfo{author}{\bibfnamefont{X.-L.} \bibnamefont{Qi}},
  \bibinfo{author}{\bibfnamefont{S.-C.} \bibnamefont{Zhang}}, \bibnamefont{and}
  \bibinfo{author}{\bibfnamefont{N.}~\bibnamefont{Nagaosa}},
  \bibinfo{journal}{Phys. Rev. Lett.} \textbf{\bibinfo{volume}{102}},
  \bibinfo{pages}{256403} (\bibinfo{year}{2009}).

\bibitem[{\citenamefont{{K\"onig} et~al.}(2007)\citenamefont{{K\"onig},
  Wiedmann, {Br\"{u}ne}, Roth, Buhmann, Molenkamp, Qi, and
  Zhang}}]{koenig2007b}
\bibinfo{author}{\bibfnamefont{M.}~\bibnamefont{{K\"onig}}},
  \bibinfo{author}{\bibfnamefont{S.}~\bibnamefont{Wiedmann}},
  \bibinfo{author}{\bibfnamefont{C.}~\bibnamefont{{Br\"{u}ne}}},
  \bibinfo{author}{\bibfnamefont{A.}~\bibnamefont{Roth}},
  \bibinfo{author}{\bibfnamefont{H.}~\bibnamefont{Buhmann}},
  \bibinfo{author}{\bibfnamefont{L.~W.} \bibnamefont{Molenkamp}},
  \bibinfo{author}{\bibfnamefont{X.-L.} \bibnamefont{Qi}}, \bibnamefont{and}
  \bibinfo{author}{\bibfnamefont{S.-C.} \bibnamefont{Zhang}},
  \bibinfo{journal}{Science} \textbf{\bibinfo{volume}{318}},
  \bibinfo{pages}{766} (\bibinfo{year}{2007}).

\bibitem[{\citenamefont{Roth et~al.}(2009)\citenamefont{Roth, {Br\"{u}ne},
  Buhmann, Molenkamp, Maciejko, Qi, and Zhang}}]{Roth2009}
\bibinfo{author}{\bibfnamefont{A.}~\bibnamefont{Roth}},
  \bibinfo{author}{\bibfnamefont{C.}~\bibnamefont{{Br\"{u}ne}}},
  \bibinfo{author}{\bibfnamefont{H.}~\bibnamefont{Buhmann}},
  \bibinfo{author}{\bibfnamefont{L.~W.} \bibnamefont{Molenkamp}},
  \bibinfo{author}{\bibfnamefont{J.}~\bibnamefont{Maciejko}},
  \bibinfo{author}{\bibfnamefont{X.-L.} \bibnamefont{Qi}}, \bibnamefont{and}
  \bibinfo{author}{\bibfnamefont{S.-C.} \bibnamefont{Zhang}},
  \bibinfo{journal}{Science} \textbf{\bibinfo{volume}{325}},
  \bibinfo{pages}{294} (\bibinfo{year}{2009}).

\bibitem[{\citenamefont{{B\"{u}ttiker}}(2009)}]{Buttiker2009}
\bibinfo{author}{\bibfnamefont{M.}~\bibnamefont{{B\"{u}ttiker}}},
  \bibinfo{journal}{Science} \textbf{\bibinfo{volume}{325}},
  \bibinfo{pages}{278} (\bibinfo{year}{2009}).

\bibitem[{\citenamefont{Kane and Mele}(2005{\natexlab{b}})}]{kane2005a}
\bibinfo{author}{\bibfnamefont{C.~L.} \bibnamefont{Kane}} \bibnamefont{and}
  \bibinfo{author}{\bibfnamefont{E.~J.} \bibnamefont{Mele}},
  \bibinfo{journal}{Phys. Rev. Lett.} \textbf{\bibinfo{volume}{95}},
  \bibinfo{pages}{146802} (\bibinfo{year}{2005}{\natexlab{b}}).

\bibitem[{\citenamefont{Wu et~al.}(2006)\citenamefont{Wu, Bernevig, and
  Zhang}}]{wu2006}
\bibinfo{author}{\bibfnamefont{C.}~\bibnamefont{Wu}},
  \bibinfo{author}{\bibfnamefont{B.~A.} \bibnamefont{Bernevig}},
  \bibnamefont{and} \bibinfo{author}{\bibfnamefont{S.~C.} \bibnamefont{Zhang}},
  \bibinfo{journal}{Phys. Rev. Lett.} \textbf{\bibinfo{volume}{96}},
  \bibinfo{pages}{106401} (\bibinfo{year}{2006}).

\bibitem[{\citenamefont{Xu and Moore}(2006)}]{xu2006}
\bibinfo{author}{\bibfnamefont{C.}~\bibnamefont{Xu}} \bibnamefont{and}
  \bibinfo{author}{\bibfnamefont{J.~E.} \bibnamefont{Moore}},
  \bibinfo{journal}{Phys. Rev. B} \textbf{\bibinfo{volume}{73}},
  \bibinfo{pages}{045322} (\bibinfo{year}{2006}).

\bibitem[{\citenamefont{Stone and Szafer}(1988)}]{stone1988}
\bibinfo{author}{\bibfnamefont{A.~D.} \bibnamefont{Stone}} \bibnamefont{and}
  \bibinfo{author}{\bibfnamefont{A.}~\bibnamefont{Szafer}},
  \bibinfo{journal}{IBM J. Res. Dev.} \textbf{\bibinfo{volume}{32}},
  \bibinfo{pages}{384} (\bibinfo{year}{1988}).

\bibitem[{\citenamefont{Slonczewski}(1989)}]{Slonczewski1989}
\bibinfo{author}{\bibfnamefont{J.~C.} \bibnamefont{Slonczewski}},
  \bibinfo{journal}{Phys. Rev. B} \textbf{\bibinfo{volume}{39}},
  \bibinfo{pages}{6995} (\bibinfo{year}{1989}).

\bibitem[{\citenamefont{Datta and Das}(1990)}]{Datta1990}
\bibinfo{author}{\bibfnamefont{S.}~\bibnamefont{Datta}} \bibnamefont{and}
  \bibinfo{author}{\bibfnamefont{B.}~\bibnamefont{Das}},
  \bibinfo{journal}{Appl. Phys. Lett.} \textbf{\bibinfo{volume}{56}},
  \bibinfo{pages}{665} (\bibinfo{year}{1990}).

\bibitem[{\citenamefont{{K\"{o}nig} et~al.}(2008)\citenamefont{{K\"{o}nig},
  Buhmann, Molenkamp, Hughes, Liu, Qi, and Zhang}}]{Konig2008}
\bibinfo{author}{\bibfnamefont{M.}~\bibnamefont{{K\"{o}nig}}},
  \bibinfo{author}{\bibfnamefont{H.}~\bibnamefont{Buhmann}},
  \bibinfo{author}{\bibfnamefont{L.~W.} \bibnamefont{Molenkamp}},
  \bibinfo{author}{\bibfnamefont{T.}~\bibnamefont{Hughes}},
  \bibinfo{author}{\bibfnamefont{C.-X.} \bibnamefont{Liu}},
  \bibinfo{author}{\bibfnamefont{X.-L.} \bibnamefont{Qi}}, \bibnamefont{and}
  \bibinfo{author}{\bibfnamefont{S.-C.} \bibnamefont{Zhang}},
  \bibinfo{journal}{J. Phys. Soc. Jpn} \textbf{\bibinfo{volume}{77}},
  \bibinfo{pages}{031007} (\bibinfo{year}{2008}).

\bibitem[{\citenamefont{Ferry and Goodnick}(1997)}]{FerryGoodnick}
\bibinfo{author}{\bibfnamefont{D.~K.} \bibnamefont{Ferry}} \bibnamefont{and}
  \bibinfo{author}{\bibfnamefont{S.~M.} \bibnamefont{Goodnick}},
  \emph{\bibinfo{title}{Transport in Nanostructures}}
  (\bibinfo{publisher}{Cambridge University Press},
  \bibinfo{address}{Cambridge}, \bibinfo{year}{1997}).

\bibitem[{\citenamefont{Wu et~al.}(1994)\citenamefont{Wu, Cocks, and
  Jayanthi}}]{Wu1994}
\bibinfo{author}{\bibfnamefont{S.~Y.} \bibnamefont{Wu}},
  \bibinfo{author}{\bibfnamefont{J.}~\bibnamefont{Cocks}}, \bibnamefont{and}
  \bibinfo{author}{\bibfnamefont{C.~S.} \bibnamefont{Jayanthi}},
  \bibinfo{journal}{Phys. Rev. B} \textbf{\bibinfo{volume}{49}},
  \bibinfo{pages}{7957} (\bibinfo{year}{1994}).

\bibitem[{\citenamefont{Sanvito et~al.}(1999)\citenamefont{Sanvito, Lambert,
  Jefferson, and Bratkovsky}}]{Sanvito1999}
\bibinfo{author}{\bibfnamefont{S.}~\bibnamefont{Sanvito}},
  \bibinfo{author}{\bibfnamefont{C.~J.} \bibnamefont{Lambert}},
  \bibinfo{author}{\bibfnamefont{J.~H.} \bibnamefont{Jefferson}},
  \bibnamefont{and} \bibinfo{author}{\bibfnamefont{A.~M.}
  \bibnamefont{Bratkovsky}}, \bibinfo{journal}{Phys. Rev. B}
  \textbf{\bibinfo{volume}{59}}, \bibinfo{pages}{{11 }936}
  (\bibinfo{year}{1999}).

\bibitem[{\citenamefont{Zhou et~al.}(2008)\citenamefont{Zhou, Lu, Chu, Shen,
  and Niu}}]{Zhou2008}
\bibinfo{author}{\bibfnamefont{B.}~\bibnamefont{Zhou}},
  \bibinfo{author}{\bibfnamefont{H.-Z.} \bibnamefont{Lu}},
  \bibinfo{author}{\bibfnamefont{R.-L.} \bibnamefont{Chu}},
  \bibinfo{author}{\bibfnamefont{S.-Q.} \bibnamefont{Shen}}, \bibnamefont{and}
  \bibinfo{author}{\bibfnamefont{Q.}~\bibnamefont{Niu}},
  \bibinfo{journal}{Phys. Rev. Lett.} \textbf{\bibinfo{volume}{101}},
  \bibinfo{pages}{246807} (\bibinfo{year}{2008}).

\bibitem[{\citenamefont{Usaj}()}]{usaj2009}
\bibinfo{author}{\bibfnamefont{G.}~\bibnamefont{Usaj}},
  \bibinfo{howpublished}{e-print arXiv:0906.4349 (2009)}.

\end{thebibliography}

%\begin{addendum}
%\item[Acknowledgements] We thank (E.-A. Kim?), T. L. Hughes,
% C.-X. Liu, M. K\"{o}nig, B. Huard, Y. Oreg, S. Raghu, C.-C. Chen and H. Yao for
% insightful discussions. We gratefully acknowledge financial
% support by ..., the National Science and Engineering Research
% Council (NSERC) of Canada, the Fonds qu\'{e}b\'{e}cois de la recherche
% sur la nature et les technologies (FQRNT), and the Stanford Graduate Program
% (SGF).
%\item[Competing Interests] The authors declare that they have no
%    competing financial interests.
%\item[Correspondence] Correspondence and requests for materials
%    should be addressed to ... (email: ...).
%\end{addendum}

\end{document}